\pdfoutput=1
\documentclass[prd,twocolumn,groupaddress,preprintnumbers,nofootinbib]{revtex4}
\usepackage{graphicx}
\usepackage{epsfig}
\usepackage{bm}
\usepackage{latexsym,amssymb,amsmath,amsfonts,amssymb,txfonts,wasysym,float}
\usepackage{scrextend}
\usepackage{mathrsfs}
\usepackage{color}

\usepackage{enumitem}

\newcommand{\postscript}[2]{\setlength{\epsfxsize}{#2\hsize}
   \centerline{\epsfbox{#1}}}

\usepackage[usenames,dvipsnames]{xcolor}
\definecolor{orange}{cmyk}{0,0.5,1,0}
\definecolor{rossoCP3}{cmyk}{0,.88,.77,.40}
\definecolor{graa}{rgb}{0.8,0.8,0.8}
\definecolor{blaa}{rgb}{0.2,0.2,0.6}

\begin{document}

\title{\color{rossoCP3}  Deciphering the Archeological Record: Further Evidence for Ultra-High-Energy
  Cosmic Ray Acceleration in~Starburst-Driven Superwinds}

\author{Luis Alfredo Anchordoqui}

\affiliation{Department of Physics and Astronomy,  Lehman College, City University of
  New York, NY 10468, USA,
}

\affiliation{Department of Physics,
 Graduate Center, City University
  of New York,  NY 10016, USA,
}

\affiliation{Department of Astrophysics,
 American Museum of Natural History, NY
 10024, USA
}

\begin{abstract}
  \vskip 2mm \noindent Very recently, the Pierre Auger and Telescope
  Array collaborations reported strong evidence for a correlation
  between the highest energy cosmic rays and nearby starburst
  galaxies, with a global significance post-trial of
  $4.7\sigma$. It is well known that the collective effect of supernovae and winds from
  massive stars in the central region of these galaxies drives a galactic-scale
  superwind
  that can shock heat and accelerate ambient interstellar or
  circumgalactic gas. In previous work we showed that, for reasonable
  source parameters, starburst-driven superwinds can be the carriers of
  ultra-high-energy cosmic ray acceleration. In this paper we assess
  the extent to which one can approach the archaeological ``inverse''
  problem of deciphering properties of superwind evolution from 
 present-day IR emission of their host galaxies. We show that the
 Outer Limits galaxy NGC 891 could provide ``smoking gun evidence'' for
  the starburst-driven superwind
  model of ultra-high-energy cosmic rays.
\end{abstract}
\date{October 2022}
\maketitle

\section{Introduction}
\label{sec:1}

It is difficult to overstate the scientific importance of
understanding the origin of the highest energy cosmic
rays. Ultra-high-energy cosmic rays (UHECRs) have been observed since
the early 60's~\cite{Linsley:1963km} and a plethora of models have been proposed to explain
their origin, including acceleration (bottom-up) processes in power
astrophysical environments and (top-down) decay of super-heavy (GUT
scale) particles~\cite{Anchordoqui:2018qom}.

In the late 90's we proposed a model
predicting that UHECR nuclei could be 
accelerated in the large-scale terminal shock of the superwinds that
flow from the central region of starburst
galaxies~\cite{Anchordoqui:1999cu}.  
A few years later we also predicted that the
highest energy cosmic ray nuclei will generate 20 to 30 degrees hot-spots around the
starbursts, because of their magnetic deflection in the Galactic
magnetic field~\cite{Anchordoqui:2002dj}.   

Over the last decade, mounting evidence has been accumulating
suggesting that the UHECR composition becomes dominated by nuclei at
high-energy end of the spectrum~\cite{PierreAuger:2010ymv,PierreAuger:2014sui,PierreAuger:2014gko,PierreAuger:2016qzj,PierreAuger:2017tlx,Watson:2021rfb}. Concurrently, the Pierre Auger
Collaboration has provided a compelling indication for a possible
correlation between the arrival directions of  
cosmic rays with energy $E \agt 10^{10.6}~{\rm GeV}$ and a model based
on a catalog of bright starburst galaxies~\cite{PierreAuger:2018qvk,PierreAuger:2022axr}. The post-trial chance
probability in an isotropic cosmic ray sky gives a Gaussian
significance of $4.0\sigma$. When data from the Telescope Array are included in the statistical analysis the
correlation with starburst galaxies is stronger than the
Auger-only result, with a post-trial significance of
$4.7\sigma$~\cite{PierreAuger:2023zxh}. In the best-fit model, $(12.1\pm 4.5)$\%
of the UHECR flux originates from the starbursts 
and undergoes angular diffusion on a von Mises-Fisher scale $\psi \sim
{15.1^{+4.6}_{-3.0}}^\circ$.\footnote{It is important to note that a likelihood analysis considering the biases induced by the coherent
deflection in the Galactic magnetic field gives best-fit parameters
which are 
consistent within 95\%CL with those obtained adopting angular
diffusion in the  
 isotropic-scattering approximation~\cite{Higuchi:2022xiv}.} 

Together these observations lead to two critical questions:
\begin{itemize}[noitemsep,topsep=0pt]
\item What imprints may the evolution of starburst-driven superwinds leave in present-day observables?
\item To what extent can we decipher this archaeological record,
  exploiting information about the present-day universe in order to
  learn about or constrain the possible acceleration of UHECR nuclei
  in starburst-driven superwinds? 
\end{itemize}
These are certainly very broad
questions, and in this paper we attempt to take a first step towards answering these questions.

Before proceeding, we pause to note that median deflections of particles in the Galactic magnetic field are estimated to be
\begin{equation}
  \theta_G \sim 3^\circ \ Z \ \left(\frac{E}{10^{11}~{\rm GeV}} \right) \,,
\end{equation}
where $Z$ is the charge of the UHECR in units of the proton charge~\cite{IceCube:2015afa}. Thus, the requirement $\theta_G \alt \psi$ implies that UHECRs
contributing to the starburst anisotropy signal should have $Z
\alt 10$ and $E/Z \sim 10^{10}~{\rm GeV}$.

\section{Starburst  Archeology}
\label{sec:2}

Starburst-driven superwinds are complex, multi-phase phenomena 
primarily powered
by the momentum and energy injected by massive stars in the
form of supernova (SN) explosions, stellar winds, and
radiation~\cite{Heckman}. According to the book, these superwinds are
ubiquitous in galaxies where the star-formation rate per unit area
exceeds $10^{-1} M_\odot \ {\rm yr}^{-1} \ {\rm kpc}^{-2}$. This type of
starbursting object, nicknamed far-IR galaxy (FIRG), can be
characterized by:  
{\it (i)}~an IR luminosity, $L_{\rm IR} \agt
10^{44}~{\rm erg \, s^{-1}}$, which is large relative to its optical
luminosity $L_{\rm IR} \gg L_{\rm OPT}$,  and {\it (ii)}~a ``warm'' IR
spectrum (flux density at $60~\mu{\rm m}> 50\%$ the flux density at
$100~\mu {\rm m}$).

The
deposition of mechanical energy by supernovae and stellar winds
results in a bubble filled with hot ($T \alt 10^8~{\rm K}$) gas that
is unbound by the gravitational potential because its temperature is
greater than the local escape temperature. The over-pressured bubble
expands adiabatically, becomes supersonic at the edge of the starburst
region, and eventually blows out of the disk into the halo forming a
strong shock front on the contact surface with the cold gas in the
halo.

Two distinct mechanisms have been proposed to explain the starburst anisotropy signal:
\begin{itemize}[noitemsep,topsep=0pt]
\item UHECRs can be accelerated by
bouncing back and forth across the superwind's terminal shock
(hereafter ARC model)~\cite{Anchordoqui:1999cu}.
\item UHECR acceleration can occur in the disproportionally frequent
  extreme explosions that take place in the
  starburst nucleus due to the high star-formation
  activity~\cite{AlvesBatista:2019tlv};  e.g., low-luminosity gamma-ray bursts (llGRBs)~\cite{Zhang:2017moz}. 
\end{itemize}
A point worth noting at this juncture is that one would expect llGRB
explosions to stochastically sample the
locations of cosmic star-formation throughout the volume of the
Universe in which they can be observed. Then the probability for a given
type of galaxy to host a llGRB during some period of time would be proportional to
its star-formation rate. Starburst galaxies represent about 1\% of the fraction
   of galaxies containing star forming galaxies~\cite{Bergvall}, and
   the probability of SN explosions is about one to two orders of
   magnitude larger in starbursts than in normal galaxies, e.g., the
   SN rate for M82 is about $0.2-0.3~{\rm yr}^{-1}$~\cite{Ulvestad:1995yt} whereas for the Milky
   Way is $\sim 3.5 \pm 1.5~{\rm century}^{-1}$~\cite{Dragicevich}. Note that these two effects tend
   to compensate each other, and so a straightforward calculation
   shows that UHECRs accelerated in llGRBs will have a stronger 
   correlation with the nearby matter distribution than with starbursts~\cite{Anchordoqui:2019ncn}.  
   
Indeed, given the ubiquity of llGRB explosions  we
   can ask ourselves why the correlation of UHECRs with starburst
   galaxies would be explained by the presence of this {\it common}
   phenomenon. Rather there must be some other inherently unique
   feature of starburst galaxies to account for this
   correlation. With this in mind, herein we focus on the ARC model~\cite{Anchordoqui:1999cu}. In previous work~\cite{Anchordoqui:2018vji,Anchordoqui:2020otc}, we investigated the constraints imposed by the
starburst anisotropy signal on the ARC model and we 
readjusted free parameters to remain consistent with the most recent
astrophysical observations. We now investigate the
minimum power requirement for UHECR acceleration at shocks.

The cosmic ray maximum energy for any multiplicative acceleration process is given by the Hillas criterion, which yields $E_{\rm max}
\sim Ze u B R$, where $R$
 is the size and  
$B$ the magnetic field strength of the acceleration region, and $u$ is
the speed of the scattering centers (i.e., the shock
velocity)~\cite{Hillas:1984ijl}. Now, the magnetic field $B$ carries
with it an energy density $B^2 /(2\mu_0)$, and the out-flowing plasma carries with
it an energy flux $\sim u R^2B^2/(2\mu_0)$, where $\mu_0$ is the permeability of free space. This sets a constraint on 
the maximum magnetic power delivered through the shock~\cite{Waxman:1995vg}. Following~\cite{Matthews:2018laz}, we combine the Hillas
criterion with the constraint of the magnetic energy flux to arrive at the minimum power needed to accelerate a nucleus to a
given rigidity ${\cal R}$,
\begin{equation}
  {\cal P}_{\rm min} = \frac{{\cal R}^2}{2 \mu_0 \ u} \sim  10^{44}~{\rm erg \,
    s^{-1}} \left(\frac{u}{0.01 \, c} \right)^{-1} \ \left(\frac{{\cal R}}{10^{10}~{\rm GV}}
  \right)^2 \, .
\label{Power}
\end{equation}
This steady state argument provides a conservative upper limit for
the required minimum power in the superwind. Note that the minimum power requirement can
be relaxed if, e.g., the energy carried by the out-flowing plasma needed to
maintain a $100~\mu{\rm G}$ magnetic field strength on a scale of 15~kpc~\cite{Anchordoqui:2020otc}  is supplied during
periodic flaring intervals. Throughout we remain cautious and adopt (\ref{Power}) as our point of reference.

Next, in line with our stated plan, we adopt the functional form
of the energy injection
rate from stellar winds and supernovae estimated in~\cite{Heckman:1990fe} to determine
the kinetic energy output of the starburst from measurements of the IR
luminosity,
\begin{equation}
  {\cal P}_{\rm today} \sim 4 \times 10^{43} \ L_{\rm IR,11}~{\rm erg} \ {\rm s}^{-1} \,,
\label{Powert}
\end{equation}  
where $L_{\rm IR,11}$ is the total IR luminosity (in units of $10^{11} L_\odot$),
and where we have rescaled the normalization factor to accommodate a  
supernova rate in M82 of $0.3~{\rm yr}^{-1}$~\cite{Bregman}, rather than the
$0.07~{\rm yr}^{-1}$ used in the
original calculation of~\cite{Heckman:1990fe}. The associated
mass-loss rate to match the normalization $u \sim 0.01 \, c$ is found to be
\begin{equation}
  \dot M \sim 15 \  L_{\rm IR,11} \ M_\odot~{\rm yr}^{-1} \ .
 \label{Mdot} 
\end{equation}  
In Table~\ref{tabla:1}
we list the present-day kinetic energy output of the nearby starbursts
contributing to the anisotropy signal.

\begin{table}
  \caption{Infrared luminosities~\cite{Sanders:2003ms} and kinetic
    energy
    output. \label{tabla:1}}
\begin{tabular}{lcc}
  \hline
  \hline
  Starburst Galaxy & ${\rm log}_{10} (L_{\rm IR}/L_\odot)$ &
                                                             ~~~~${\cal P}_{\rm today}/(10^{43} {\rm erg \, s^{-1}})$~~~~ \\
  \hline
NGC 253     &     $10.44$ & 1 \\
  NGC 891     &     $10.27$ & 0.7 \\
NGC 1068    &     $11.27$ & 7 \\
NGC 3034 (a.k.a. M82) ~~~~ & $10.77$ & 2\\
NGC 4945 & $10.48$ & 1  \\
NGC 5236 (a.k.a. M83)  & $10.10$ & 0.5 \\
NGC 6946 & $10.16$ & 0.6 \\
  IC 342 & $10.17$ & 0.6 \\
  \hline
  \hline
\end{tabular}
\end{table}

By comparing (\ref{Power}) with the results on Table~\ref{tabla:1} we
see that for most of the starbursts the present-day power output falls
short by about an order of magnitude to accommodate the required
maximum rigidity to explain the anisotropy signal. However, we note
that the estimate in (\ref{Power}) is subject to large systematic
uncertainties; see Appendix I. Furthermore, the characteristic
time-scale for Fermi-acceleration in non-relativistic shocks is
${\cal O} (10^7 {\rm yr})$~\cite{Fermi:1954ofk}, and the superwind
power given in Table~\ref{tabla:1} does not take into account any
source evolution, but rather characterizes the current state of the
outgoing plasma assuming that the star formation proceeded continuously at a constant
rate.

The question is then: Could the superwind of the FIRGs listed in
Table~\ref{tabla:1} be more powerful in an earlier stage? The answer
to this question is, in principle, yes: the rationale being that very
powerful FIRGs have been observed in our cosmic backyard. For example,
Arp 220 and NGC 6240 are the nearest and best-studied examples of very
powerful FIRGs ($L_{\rm IR} \sim 10^{12} L_\odot$), while IRAS 00182 -
7112 is the most FIR-powerful galaxy yet discovered ($L_{\rm IR}$
nearly $10^{13} L_\odot$ )~\cite{Heckman:1990fe}. We note, however,
that there is no solid evidence indicating that these powerful FIRGs
could represent an early stage of the starburst evolution. The star
formation history of Arp 220 is multiplex: in the central kpc, a 10~Myr
old starburst provides the majority of the IR
luminosity~\cite{Engel:2011zw}. This is comparable to the mean age of
the active starburst in M82, but somewhat younger than the stellar
activity powering the central regions of NGC 253, see Appendix I. What's
more, data from supernova radio spectra of Arp 220 may be indicative
of a very short, intense burst of star formation around $3~{\rm Myr}$
ago~\cite{Parra:2006mm}, which could be the elephant in the room.

Now, the anisotropy correlation is observed to be spread over an
angular region of size $\sim 20^\circ$.\footnote{The angular spread of
  a von Mises-Fisher distribution, the equivalent of a Gaussian on the
  sphere, corresponds to a top-hat scale
  $\Psi \sim 1.59 \times \psi$.} The spread originates in the
scattering of UHECRs with magnetic fields as they propagate to the
Earth. The scattering on magnetic fields also gives rise to a spread
in the arrival times of UHECRs with respect to the IR emission. If all
scattering takes place in the Milky Way the dispersion in the UHECR
arrival times is roughly $10^3~{\rm yr}$~\cite{Pfeffer:2015idq}. This
seems a very short period of time to accommodate an order of magnitude
change in ${\cal P}$. However, since the  number density of gas in the
halo region is $10^{-3} \alt n_{\rm g,h}/{\rm cm}^{3} \alt 10^{-2}$~\cite{Romero:2018mnb,Tomisaka:1992vq,Strickland:2000jg,Sharma:2013kza} and the total $pp$ cross section
at $E \sim 10^{10}~{\rm GeV}$ is $\sigma_{pp} \sim 100~{\rm mb}$, the
characteristic time scale for $pp$ collisions in the acceleration
region is $1 \alt \tau^{pp}_{\rm int}/{\rm Gyr} \alt 10$. For a nucleus
of baryon number $A$ scattering off a proton, an order of magnitude
estimate of the total cross section can be obtained from the empirical
scaling $\sigma_{Ap} \sim A^{2/3} \sigma_{pp}$~\cite{Abu-Ibrahim:2010vii}. Thereby,  for a medium
mass nucleus (carbon, nitrogen, or oxygen), the characteristic time
scale to scatter off the gas in the acceleration
region is $0.1 \alt \tau^{A p}_{\rm int}/{\rm Gyr} \alt 1$. Moreover, the energy
loss rate of baryon scattering is
comparable to that of nucleus photodisintegration in the background
radiation fields of the acceleration region, see Appendix II. All in all, this implies
that even considering the upper range for estimates of the gas density
in the halo, UHECRs ARC accelerated by the starburst superwind could be storaged in the acceleration region without suffering catastrophic
spallations.

At last, we adopt the venerable leaky-box approximation to outline a first order
model that can describe the dynamics of the starburst emission process. In
analogy with the {\it closed Galactic model} of~\cite{Rasmussen:1975kg}, we start with the assumption
that, in a first stage, during the ARC acceleration process all cosmic rays
remain confined to the galaxy by a suitable large magnetic field, even
at the highest energies. In a second stage, the confinement power of
the magnetic field (correlated with $L_{\rm IR}$) gradually decreases
and the UHECRs are trapped, but within reflecting boundaries
surrounding the galaxy, such that at each encounter with the boundary,
they have a {\it time-dependent} probability of escaping into the intergalactic
space. This leakage of UHECRs is driven by diffusion with a time-dependent rigidity,
but of course the larger the
rigidity the less time spent in the confinement volume. 

To develop some sense for the orders of magnitude involved, we recall
that in each ``scatter'', the diffusion coefficient describes an
independent angular deviation of particle trajectories
whose magnitude depends on the Larmor radius $r_L = 0.1 Z^{-1} \,
E_{10} \, B^{-1}_{100}~{\rm kpc}$, where $E_{10^{10}} = E/10^{10}~{\rm
  GeV}$ and $B_{\rm 100} = B/100~\mu{\rm
  G}$~\cite{Anchordoqui:2018qom}.\footnote{Our fiducial choice of $B$ is
  supported by full-blown simulations, which accurately capture the
  hydrodynamic mixing and dynamical interactions between the hot and
  cold phases in the out-flow~\cite{Buckman:2019pcj}; for details, see~\cite{Anchordoqui:2020otc}.} That said, it is reasonable to assume that
a medium mass nucleus with $E \alt 10^{10.7}~{\rm GeV}$ would remain trapped inside
magnetic subdomains of size $\ell \sim 0.1~{\rm kpc}$, attaining
efficient diffusion when the wave number of the associated Alfv\'en
wave is equal to the Larmor radius of the nucleus~\cite{Wentzel:1974cp}. With a Kolmogorov form for the turbulent
magnetic field power spectrum (with coherent directions
on scales $\ell$), this gives for a diffusion coefficient
\begin{equation}
  D(E) \sim 5 \ \ell_{0.1}^{2/3} \ B_{100}^{-1/3} \
  E^{1/3}_{10^{10}}~~\frac{\rm kpc^2}{\rm Myr} \,,
\end{equation}
where $\ell_{0.1} = \ell/0.1~{\rm
  kpc}$~\cite{Blasi:1998xp}. Neglecting a numerical constant of order
one, the diffusion time to the ``magnetic
wall'' is estimated to be $\tau_{\rm diff} (E) \sim H^2/D(E)$, where $H$ is the
height of the halo~\cite{Gaisser:1990vg}. For $H \sim 15~{\rm
  kpc}$~\cite{Strickland:2001te}, the diffusion time is in the
ballpark, $\tau_{\rm diff} (E) \sim 50~{\rm
  Myr}/E_{10^{10}}^{1/3}$. Note that a mean residence time in the
confinement volume of $\tau_{\rm
  res} \sim 0.1~{\rm Gyr}$, would require cosmic rays to bounce one ($E
\sim 10^{10}~{\rm GeV}$) or two ($E \sim 10^{10.7}~{\rm GeV}$) times
in the reflecting boundaries. Therefore, for the problem at hand, the
ratio of the diffusion coefficient in the wall to that in the
confinement volume must not be unreasonably high. This is in sharp
contrast to the classical leaky box description of Galactic cosmic
rays~\cite{Jones:1990}.  

Needless to say, the {\it ad hoc} assumption that the residence time
of UHECRs is compatible with a single value is a rudimentary order of magnitude
approximation, because particles originated in the center reside longer
that those populating the halo
periphery. A precise description of the storage/trapping mechanism,
with a {\it time-dependent} probability of escaping into the
intergalactic space, requires a numerical simulation that is beyond the scope of
this paper. Besides, one may object our choice of fiducial
values for $B$ and $H$, which could be considered somewhat
off-center. Whichever point of view one may find more convincing, it seems most
conservative at this point to depend on experiment (if possible) to
resolve the issue. In the next section we examine a particular FIRG example
that could provide a clear-cut experimental test of the ideas discussed above.

\section{NGC 891: The smoking gun}
\label{sec:3}

The Outer Limits spiral galaxy NGC 891 lives in the Local Super Cluster,
some 9.7~Mpc away.\footnote{NGC 891 appears in the end credits of the
  Outer Limits 1963 TV series.} The galaxy looks as the Milky Way would
look like when viewed edge-on, and indeed the two galaxies are
considered very similar in terms of luminosity and
size~\cite{Karachentsev:2004dx}. However, HI observations reveal
structures on various scales. High resolution images of its dusty disk
show unusual filament patterns extending into the halo of the galaxy,
away from its galactic disk. The overall kinematics of the gas in the
halo is characterized by differential rotation lagging with respect to
that of the disk. The lag, which is more pronounced at small radii,
increases with height from the plane. There is evidence that a
significant fraction of the halo is due to a galactic
fountain~\cite{Oosterloo:2007se}.

The slope of the luminosity function is powerful indicator of the main
features of the host galaxy. On the one hand, if the star formation rate in the
galaxy is constant, then the luminosity function describing the sources
can be accommodated by a single unbroken power-law model. On the other
hand, if the galaxy is starbursting  then new high-mass X-ray binaries
would be formed, breaking the luminosity function slope. The break in
the slope would decrease with time, and would be an indication of the
time of previous bursts in the galaxy. The X-ray luminosity
function of NGC 891 can be fitted by a single power-law, but with a
slope typical of starburst galaxies and 
flux density ratio (at $60~\mu{\rm m}/100~\mu{\rm m}$) corresponding more to normal spirals than
starbursts~\cite{Temple:2005sx}. Altogether this indicates that NGC 891
is a starburst in a {\it quiescent state}.

NGC 891, located in equatorial coordinates at (right ascension,
declination) = $(35.64^\circ, 42.35^\circ)$, is inside the $\sim
20^\circ$ angular window where the Telescope Array 
Collaboration reported an excess of events over expectations from isotropy~\cite{TelescopeArray:2021dfb}.  More precisely, the
hot-spot is centered at $(19^\circ, 35^\circ)$, and has a local
significance of $4\sigma$ down to
$E \sim 10^{10.1}~{\rm GeV}$.\footnote{We used $E_{\rm Auger} = E_0
  e^\alpha (E_{\rm TA}/E_0)^\beta$, with $E_0 = 10^{10}~{\rm GeV}$,
  $\alpha = - 0.159 \pm 0.012$ and $\beta = 0.945 \pm
  0.016$~\cite{PierreAuger:2023zxh}.} More data are certainly needed to ascertain
whether the hot spot originates in NGC 891. Should this be the case,
it is clear that UHECR acceleration must have occurred in the past, when
the galaxy was starbursting. 

Note that, in very good approximation, UHECRs are emitted instantaneously by llGRB and they cannot remain unscathed inside the starburst
core. Indeed, the production of very- and ultra-high energy neutrinos
would be a clear indication that UHECRs are accelerated in the central
region of the 
starburst~\cite{Condorelli:2022vfa,Muzio:2022bak}. In contrast,
in the ARC model, the interaction of UHECRs with the gas and radiation
fields in the halo is largely suppressed
(for details, see Appendix~III), and
therefore the neutrino signal would be largely suppressed.

We stress once more that NGC 891 is a starburst
in a {\it quiescent state}. Because of this, the production of UHECRs
assuming typical starburst properties must have occurred in the past,
when the starbursting nucleus was active. Now, since these cosmic rays
cannot be storaged in the nucleus, but only in the halo, we can
conclude that the
confirmation of NGC 891 as an UHECR source would disfavor acceleration
through powerful explosions in the starburst nucleus, while providing some support for ARC
acceleration with time-dependent reflecting boundaries surrounding the galaxy.

\section{Conclusions}
\label{sec:4}

We have investigated whether starburst-driven superwinds satisfy the
power requirements to accelerate UHECRs. Since the acceleration time
scale is ${\cal O} (10^7~{\rm yr})$,  this quest by necessity has
been fundamentally archaeological: we seek to exploit
present-day observations in the interest of learning about the
past. We have shown that NGC 891 provides fossil evidence that UHECRs 
may have been
ARC accelerated in the past and
remained storaged in the acceleration region, where they diffused
rather freely without suffering significant
energy losses. When the
confinement power of the magnetic field (correlated with $L_{\rm IR}$)
in the storage region 
started to gradually decrease a fraction of the cosmic rays began to
escape at a rate that has slowly increased over time.

\section*{Acknowledgments}

The research of LAA is supported by
the U.S. NSF grant PHY-2112527.

\section*{Appendix I: Systematic Uncertainties}

The estimate of the kinetic energy output for the starbursts is model
dependent, and therefore subject to large systematic
uncertainties. One source of uncertainty is the functional form
relating ${\cal P}$ and $L_{\rm IR}$. A commonly-used,
 simplifying assumption when modeling the starburst histories is
that the star formation proceeds continuously at a constant
rate~\cite{Leitherer:1999rq}. For a constant star formation rate (SFR), the Salpeter stellar initial
mass function (IMF)~\cite{Salpeter:1955it} leads to ratios ${\cal P}_{\rm
  today}/L_{\rm IR,11}$ and $\dot M/L_{\rm IR,11}$ which are a factor
of $\sim 3$ smaller than the fiducial values
given in (\ref{Powert}) and (\ref{Mdot})~\cite{Veilleux:2005ia}. A
factor of ${\rm SFR}_{[46]}/{\rm SFR}_{[45]} \sim 1.5$  comes from the adopted
relation between SFR and $L_{\rm IR}$ derived
in~\cite{Hunter} assuming that the mass spectrum between 10 and 100 $M_\odot$ follows a
power-law with slope 2.35~\cite{Leitherer:1998tn}. IMFs which are
considerably flatter than the Salpeter IMF lead to higher values of
${\cal P}_{\rm today}$~\cite{Elson}. However, the point
we want to make herein is not only that the uncertainty in the empirically
determination of the proportionality
constant relating ${\cal P}_{\rm today}$ and $L_{\rm IR}$ is large,
but also that for many purposes, a steady-state approximation of the
starburst phenomenon is inadequate,
and consequently the functional form relating ${\cal P}$ and $L_{\rm IR}$
must be time-dependent. In
particular, there is evidence for more than one starbursting region in
the nearby galaxy M82, with one of these star forming regions active
until $\sim 20~{\rm Myr}$ ago, but suppressed over the last 10 to
$15~{\rm Myr}$~\cite{deGrijs:2000iv,deGrijs:2001ec}. The mean age of
the active starburst is $\alt 15~{\rm Myr}$~\cite{deGrijs:2000iv}. NGC 253 has also a complex
structure, with the possibility of two or more (a)synchronous starbursting episodes during its
lifetime~\cite{Davidge:2010qg}. The central regions of NGC 253 contain
a large population of rather young stars formed within the past $\sim 30~{\rm
  Myr}$~\cite{Davidge}. 

A second source of uncertainty steams from the shock
velocity. Parameterized three-dimensional radiative hydrodynamical
simulations that follow the emergence and dynamics of starburst's superwinds seem to indicate that superwinds driven by
energy-injection in a ring-like geometry can produce a fast ($\agt 0.01 c$) transient flow along the minor axis~\cite{Nguyen}. 

\section*{Appendix II: Photodisintegration Time Scale}

\begin{figure}
  \postscript{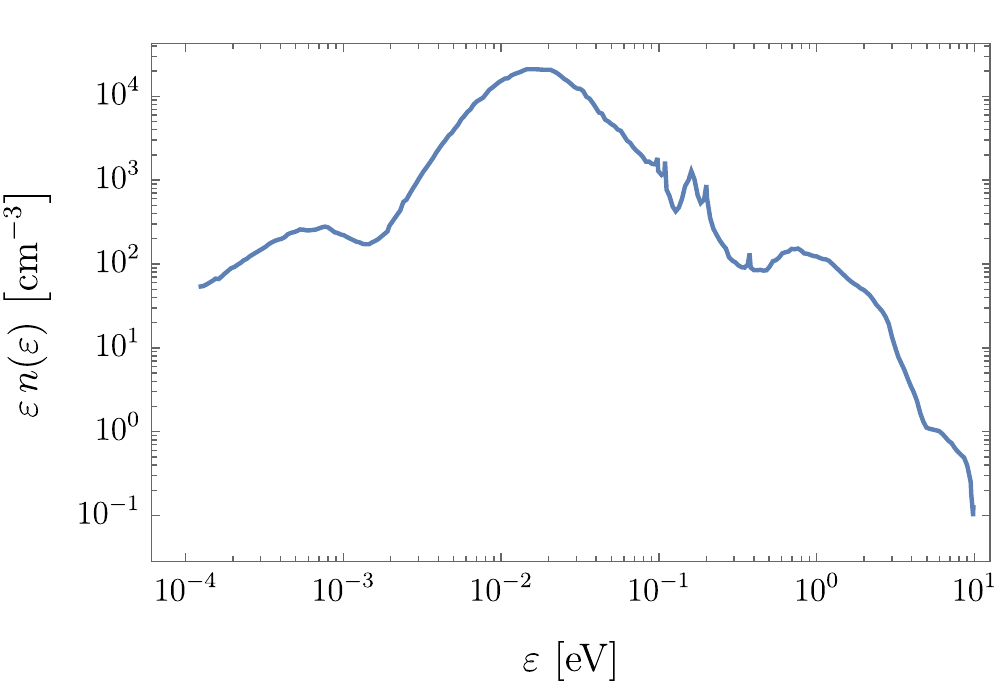}{0.8}
  \caption{Spectral energy density  for the radiation fields of M82~\cite{Lacki:2010ue}. \label{fig:1}}
  \end{figure}
In this Appendix, we estimate the energy loss rate due to UHECR
nucleus photodisintegration. We approximate the starburst core as a 
cylinder of radius $\sim 200~{\rm pc}$ and thickness $\sim 80~{\rm
  pc}$. The spectral energy density for the radiation fields of M82
inside the core is shown in
Fig.~\ref{fig:1}. It can be approximated by a broken power law,
\begin{equation}
      n(\varepsilon) = n_0
        \begin{cases}
           (\varepsilon/\varepsilon_0)^\alpha & \varepsilon < \varepsilon_0 \\
           (\varepsilon/\varepsilon_0)^\beta & \text{otherwise}  
        \end{cases} \,,
\label{eq:photonfield}
\end{equation}
where $\varepsilon$ is the photon energy, the maximum photon number density is at
an energy of $\varepsilon_0 = 11~{\rm meV}$, and where $n_0 = 10^6~{\rm
  cm}^{-3} \, {\rm eV}^{-1} $, $\alpha = 1.1$, and $\beta =
-1.8$~\cite{Muzio:2022bak}. The spectrum is normalized so that the
total number density of photons is $\int n(\varepsilon) d
\varepsilon$. Out in the halo, $|z|
\agt  40~{\rm pc}$, for $\varepsilon \agt 10^{-3}~{\rm eV}$, the photon density decreases with the
square of the distance, i.e.,
$n_0 = 10^6~{\rm
  cm}^{-3} \, {\rm eV}^{-1}  \ (z/40~{\rm pc})^{-2}$. For $\varepsilon
\alt 10^{-3}~{\rm eV}$, the spectral energy density is dominated by the
contribution of the cosmic microwave background.

The photodisintegration time scale of a cosmic ray nucleus, with energy $E
= \gamma A m_p$ and Lorentz boost $\gamma$,  propagating
through an isotropic photon background with energy $\varepsilon$ and
spectral energy density $n(\varepsilon)$ is found to be
\begin{equation}
 \frac{1}{\tau_\mathrm{int}} = \frac{c}{2} \,\int_0^\infty
                 d\varepsilon \,\frac{n(\varepsilon)}{\gamma^2 \varepsilon^2}\, \int_0^{2\gamma\varepsilon}
                 d\varepsilon^\prime \, \varepsilon^\prime\, \sigma(\varepsilon^\prime),
\label{app:eq:interaction}
\end{equation}
where $\sigma(\varepsilon^\prime)$ is the photodisintegration
cross section  by a photon of energy
$\varepsilon'$ in the rest frame of the nucleus, and where $A$ is the nucleus baryon number and $m_p$ the mass of
the proton~\cite{Stecker:1969fw}.

The photodisintegration cross section is
approximated by a single pole in the narrow-width
approximation,
\begin{equation}
\sigma (\varepsilon') = \pi\,\,\sigma_{\rm res}\,\,  \frac{\Gamma_{\rm
  res}}{2} \,\,
\delta(\varepsilon' - \varepsilon'_{\rm res})\, ,
\label{sigma}
\end{equation}
where $\sigma_{\rm res} \approx 1.45\times 10^{-27}~{\rm cm}^2 \, A$ is the resonance peak, $\Gamma_{\rm res} = 8~{\rm MeV}$ its
width, and $\epsilon'_{\rm res} = 42.65
A^{-0.21} \, (0.925 A^{2.433})~{\rm MeV},$ for $A > 4$ ($A\leq
4$) the pole~\cite{Karakula:1993he}.  The factor of $1/2$ is introduced to match the integral
(i.e. total cross section) of the Breit-Wigner and the delta
function~\cite{Anchordoqui:2006pe}.

Substitution of (\ref{sigma}) into (\ref{app:eq:interaction}) leads to,
\begin{eqnarray}
  \frac{1}{\tau_{\rm int} (E)} & \approx & \frac{c\, \pi\,
    \sigma_{\rm res}
    \,\varepsilon'_{\rm res}\,
\Gamma_{\rm res}}{4\,
    \gamma^2}
  \int_0^\infty \frac{d \varepsilon}{\varepsilon^2}\,\,\, n(\varepsilon) \,\,\,
  \Theta (2 \gamma \varepsilon - \varepsilon'_{\rm res}) \nonumber \\
  & = & \frac{c \, \pi \, \sigma_{\rm res} \,\varepsilon'_{\rm res}\,
    \Gamma_{\rm res}}{4 \gamma^2}
  \int_{\epsilon'_{\rm res}/2 \gamma}^\infty \frac{d\varepsilon}{\varepsilon^2}\,\,
  n (\varepsilon)  \, .
 \label{A1}
\end{eqnarray}
We adopt the spectral energy density of  photons in M82 as benchmark, and so substituting (\ref{eq:photonfield}) into (\ref{A1}) we arrive at
\begin{equation}
\frac{1}{\tau^{A \gamma}_{\rm int} (E)} = \frac{1}{\tau_b}
\left\{\begin{array}{ll}  \,
(E_b / E)^{\beta +1} & ~ E \leq E_b  \\
\eta (E) +
\left(E_b/E\right)^2 & ~ E > E_b
\end{array} \right. \, ,
\end{equation}
where
\begin{equation}
\tau_b = \frac{ 2 \ E_b \ (1-\beta)} {c \, \pi \
  \sigma_{\rm res} \, A \, m_p \ \Gamma_{\rm res}
  \ n_0} \,,
\end{equation}
\begin{equation}
E_b = \frac{\varepsilon'_{\rm res} \ A \ m_p}{2 \varepsilon_0} ,
\end{equation}
and
$\eta (E) = (1-\beta)/(1-\alpha) \left[\left( E_b/E \right)^{\alpha +1} -
  \left(E_b/E\right)^2 \right]$~\cite{Unger:2015laa}. A straightforward calculation shows that
 the photodisintegration time scale of medium mass (carbon,
 nitrogen, or oxygen) cosmic ray nucleus with $E \agt
10^{10}~{\rm GeV}$  propagating in the photon background at
$z \agt 1~{\rm kpc}$ is $\tau^{A\gamma}_{\rm int} \agt 0.1~{\rm Gyr}$.

Note that for a medium mass nucleus, the
energy loss per collision $\sim \gamma m_p$ is negligible when compared to
that of baryon scattering. Thus, the
energy loss rate of hadronic collisions in the halo is comparable to
that of photodisintegration.

In closing, we note that at the peak, the $\Delta (1232)$ resonant cross
section is roughly an order of magnitude smaller than the giant-dipole
resonance; for $^{16}$O, these cross sections peak at 20~MeV and 300~MeV and are roughly
30~mb~\cite{Lyutorovich:2012jd} and
3~mb~\cite{Morejon:2019pfu}, respectively). Besides, an oxygen nucleus of
$10^{10.7}~{\rm GeV}$ has a Lorentz boost $\gamma \sim 10^{9.5}$, and
so to excite the $\Delta (1232)$ would require photons with
$\varepsilon \sim 0.1~{\rm eV}$. This is a decade of energy above the
peak of the spectrum where the number of photons decreases by more
than an order of magnitude, see Fig.~\ref{fig:1}. Altogether, the interaction time scale of
photopion production is orders of magnitude larger than that of photodisintegration.

\section*{Appendix III: Assessing the Neutrino Flux}

High-energy astrophysical neutrinos are the tell-tale signature of
hadronic interactions. In this Appendix we examine whether the neutrino signal
can be used to discriminate acceleration models which take place inside 
the starburst nuclei from those in which acceleration takes place out in
the galactic halo. 

Acceleration of UHECRs in the starburst nucleus has been proposed to
originate in young neutron star
winds~\cite{Blasi:2000xm,Anchordoqui:2017abg}, tidal disruption
events~\cite{Pfeffer:2015idq,Farrar:2014yla}, and llGRB
explosions~\cite{Zhang:2017moz}. For relativistic winds of
fast-spinning pulsars, the estimated neutrino flux produced while
UHECRs cross the supernova ejecta surrounding the stars is within the
IceCube reach~\cite{Fang:2013vla,Fang:2015xhg}.

Independently of the acceleration mechanism, interactions of UHECR
within the starburst nucleus could also lead to a
measurable neutrino flux. The gas density in the starburst nucleus is
related to the star formation rate and hence associated to the IR
luminosity via the Kennicutt-Schmidt scaling~\cite{Kennicutt:1997ng},
and is given by 
\begin{equation}
n_{\rm g,n} \sim 100 \left(L_{\rm IR}/L_{\rm IR,\, M82} \right)^{0.715}~{\rm cm}^{-3} \,,  
\label{ngn}
\end{equation}
where, as noted in~\cite{Condorelli:2022vfa}, the exponent of the correlation is in agreement with~\cite{Kennicutt:2012ea}. Taking the $\sigma_{pA}$ cross section introduced in
Sec.~\ref{sec:2}, it is straightforward to see that the interaction
time-scale for UHECRs to collide with the background gas is ${\cal O}
(10^4~{\rm yr})$. On the other hand, for the radiation background shown in
Fig.~\ref{fig:1} the photodisintegration time-scale in the starburst
nucleus is ${\cal O} (0.1~{\rm Myr})$. 

In addition, the contribution of nucleus photodisintegration to the neutrino
flux originates in the decay of the emitted neutron. Note that the
maximum $\bar \nu$ energy in the neutron rest frame is very nearly
$Q_{\bar \nu} \equiv m_n - m_p - m_e = 0.78~{\rm MeV}$, where $m_n$
and $m_e$ are the mass of the neutron and electron, respectively~\cite{Anchordoqui:2003vc}. In the lab,
the ratio of the maximum $\bar \nu$ energy to the neutron energy is
$Q_{\bar \nu}/m_n \sim 10^{-3}$, and so the boosted $\bar \nu$ has an
average energy $E_{\bar \nu} \sim 10^{-3} \gamma m_n$. This is in
sharp contrast to baryon scattering, where the maximum neutrino energy
could be about 0.1 that of the incoming cosmic ray~\cite{Ahlers:2005sn}. Therefore, neutrino production is largely dominated by hadronic
collisions on the gas environment of the starburst nucleus. An
estimate of the expected diffuse neutrino flux from starbursts
assuming UHECR interactions with
the gas density given in (\ref{ngn}) has been carried out in~\cite{Condorelli:2022vfa}. For a neutrino
energy in the range $10^6 \alt  E_\nu/{\rm GeV} \alt 10^8$, the estimated single-flavor energy-squared weighted neutrino flux,
\begin{equation}
  E_\nu^2 \ \Phi_\nu \sim 10^{-8.5}~{\rm GeV} \, {\rm cm}^{-2} \, {\rm
    s}^{-1} \,  {\rm sr}^{-1} \,,   
\end{equation}
saturates the IceCube flux/limit~\cite{IceCube:2020wum}, and is therefore within reach of next
generation experiments. 

We have seen in Appendix II that out in the galactic
halo, where UHECR accelerated in the starburst superwind could be
storaged, the interaction time scale of photodisintegration and/or 
scattering off the gas is between 4 to 5 orders of magnitude
larger than the one in the nucleus. Therefore, the associated neutrino
flux would be largely suppressed. Indeed the expected neutrino flux
would be below the
cosmogenic neutrino flux and out of reach of next generation experiments; see Fig.~9 in~\cite{Condorelli:2022vfa}.

\end{document}